\documentclass[aps,prb,twocolumn,letterpaper,superscriptaddress,showpacs]{revtex4}
\usepackage{graphicx}
\usepackage{CJK}
\usepackage{amsmath,amssymb}

\usepackage{tikz}

\newcommand{\rgfao}{RbGd$_{2}$Fe$_{4}$As$_{4}$O$_{2}$}
\newcommand{\rdfao}{RbDy$_{2}$Fe$_{4}$As$_{4}$O$_{2}$}
\newcommand{\rfa}{RbFe$_{2}$As$_{2}$}
\newcommand{\kfa}{KFe$_{2}$As$_{2}$}
\newcommand{\ckfa}{CaKFe$_4$As$_4$}
\newcommand{\cfa}{CaFe$_2$As$_2$}
\newcommand{\gfao}{GdFeAsO}
\newcommand{\bkfa}{Ba$_{1-x}$K$_x$Fe$_2$As$_2$}
\newcommand{\btfao}{Ba$_2$Ti$_{2}$Fe$_{2}$As$_{4}$O}
\newcommand{\svfao}{Sr$_2$VFeAsO$_3$}

\begin{document}
\begin{CJK*}{UTF8}{bsmi}
\title{
M\"ossbauer spectroscopy study of magnetic fluctuations in superconducting RbGd$_2$Fe$_4$As$_4$O$_2$
}
\author{Yang Li}
\affiliation{Institute of Applied Magnetics, Key Lab for Magnetism and Magnetic Materials of the Ministry of Education, Lanzhou University, Lanzhou 730000, P.R. China}
\author{Zhicheng Wang}
\author{Guanghan Cao}
\affiliation{Department of Physics, Zhejiang University, Hangzhou 310027, P.R. China}
\author{Junming Zhang}
\author{Bo Zhang}
\author{Tao Wang}
\author{Hua Pang}
\author{Fashen Li}
\author{Zhiwei Li}\email{zweili@lzu.edu.cn}
\affiliation{Institute of Applied Magnetics, Key Lab for Magnetism and Magnetic Materials of the Ministry of Education, Lanzhou University, Lanzhou 730000, P.R. China}
\date{\today}

\begin{abstract}
$^{57}$Fe M\"ossbauer spectra were measured at different temperatures between 5.9\,K and 300\,K on the recently discovered self-doped superconducting \rgfao\ with T$_c$ as high as 35\,K. Singlet pattern was observed down to the lowest temperature measured in this work, indicating the absence of static magnetic order on the Fe site. The intermediate isomer shift in comparison with that of the samples \rfa\ and \gfao\ confirms the self doping induced local electronic structure change. Surprisingly, we observe two magnetic fluctuation induced spectral broadenings below $\sim$15\,K and $\sim$100\,K which are believed to be originated from the transferred magnetic fluctuations of the Gd$^{3+}$ moments and that of the magnetic fluctuations of the Fe atoms, respectively.
\end{abstract}


\maketitle
\end{CJK*}

\section{Introduction}
The discovery of superconductivity (SC) at 26\,K in LaFeAsO$_{1-x}$F$_x$ \cite{YKamihara2008jacs} has led to many following discoveries of new superconductors containing the same basic FeX (X = As, Se) units that are responsible for the high temperature SC \cite{YLSun2012jacs,AIyo2016jacs,ZCWang2016jacs,ZCWang2017jpcm,GRStewart2011rmp}. Among many other topics, tremendous results have been made regarding the phenomenon of SC and its interplay with magnetism \cite{AJDrew2009nm,PDai2012np,MDLumsden2010jpcm,DJScalapino2012rmp}. One of the most promising scenarios for the observed SC is the spin fluctuation mediated pairing mechanism which is the common thread linking a broad class of superconductors including not only the Fe and Cu based materials but also the heavy fermion superconductors  \cite{DJScalapino2012rmp}. However, the debates on this issue have never stopped and further investigation on other new families of superconductors is highly encouraged.

As for the Fe based superconductors, SC can either be induced by chemical doping or external pressure \cite{GRStewart2011rmp} and SC in both situations were found to be closely related to their magnetic properties \cite{PDai2012np,MDLumsden2010jpcm,DJScalapino2012rmp}. Interestingly, a third route towards SC, namely the self doping effect, was realized in \svfao\ \cite{GHCao2010prb}, \btfao\ and \rgfao\ \cite{YLSun2012jacs,ZCWang2017jpcm,JZMa2014prl}. Rich magnetism has been found and studied with SC for \svfao\ and it is found that magnetism and SC coexist in the freshly prepared \svfao\ \cite{XMMa2014epl}. However, the magnetism disappeared after long term storage of the sample indicating that it is meta-stable in nature which might be induced by defects or residual stress \cite{XMMa2014epl}. For the \btfao\ superconductor, neutron scattering measurements have been performed and no noticeable magnetism has been found. Additionally, nonmagnetic first principle calculations reproduced well the measured phonon spectra indicating again the nonmagnetic character of the superconducting \btfao\ \cite{MZbiri2017prb}.

On the other hand, Gd$^{3+}$ moments were suggested to order at low temperatures within the superconducting ground state in \rgfao\ \cite{ZCWang2017jpcm}. Despite the fact that the Gd$^{3+}$ moments order in separate layers than the superconducting FeAs layer, a local probe study to see if there is any transferred magnetic fluctuations from the Gd$^{3+}$ moments at the Fe site is very interesting. In addition, investigation of the local electronic structure around the Fe nucleus should be helpful to have a better understanding of the self doping induced SC.

Fruitful results regarding the local electronic structure and magnetic properties of the Fe based superconductors have already been obtained by M\"ossbauer spectroscopy in the past \cite{SIShylin2015epl, IPresniakov2013jpcm,SLBudko2016prb,ABlachowski2011prb,AOlariu2012njp,TNakamura2012jpsj,MAMcguire2009njp,ZLi2011prb}.
Therefore, in this work, we performed detailed $^{57}$Fe M\"ossbauer spectroscopy measurements on the sample \rgfao\ in the temperature range of 5.9\,K to 300\,K. Our results reveal an intermediate electronic structure between that of the Fe in \rfa\ and \gfao, suggesting that the self hole-doping effect really extends to the Fe site. More interestingly, we found evidence of magnetic fluctuations at the Fe site not only due to that transferred from the Gd$^{3+}$ moments but also due to the Fe moments itself. These results provide us a new system to further study the interplay of magnetism with SC.

\section{Experiments}

Polycrystalline material of \rgfao\ was synthesized by a solid-state reaction method. Detailed preparation procedure and its physical properties have been reported earlier \cite{ZCWang2017jpcm}. Two-phase Rietveld analysis of the X-ray diffraction pattern showed that the mass fraction of the main phase \rgfao\ amounts to $94\%$ and the only detectable impurity phase is unreacted Gd$_2$O$_3$ \cite{ZCWang2017jpcm}. As shown later, this impurity phase is transparent to our $^{57}$Fe M\"ossbauer measurements since it does not contain Fe element and thus has no influence on the data analyses. \rfa\ single crystal was grown with self flux method and measured at room temperature for comparison purpose. Detailed information with further physical characterization will be reported separately.

Transmission M\"ossbauer spectra (MS) at temperatures between 5.9\,K and 300\,K were recorded using a conventional spectrometer working in constant acceleration mode with a $\gamma$-ray source of 25\,mCi $^{57}$Co(Rh) vibrating at room temperature. The drive velocity was calibrated using sodium nitroprusside (SNP) powder and the isomer shifts (IS) quoted in this work are relative to that of the $\alpha$-Fe foil at room temperature. The full linewidth (LW) at half maximum of the SNP spectrum was 0.244(2)\,mm/s and this value can be regarded as the resolution of the spectrometer. The absorber was prepared with a surface density of $\sim$10\,mg/cm$^2$ of natural iron. All the MS were analyzed with MossWinn 4.0 \cite{mosswinn} programe.

\section{Results and discussion}
\begin{figure}
\centering
\includegraphics[width=0.9\columnwidth,clip=true]{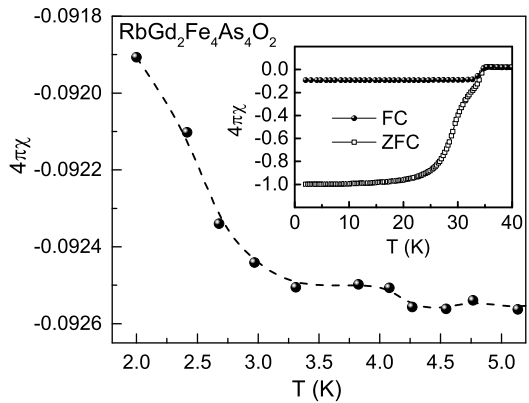}
\caption{
(color online)
Temperature dependence of the $4\pi\chi$ data of \rgfao\ measured in the FC mode with 10\,Oe magnetic field, revealing the ordering temperature of the Gd$^{3+}$ moments, T$_N\sim$3\,K. Inset shows both the FC and ZFC data, showing the superconducting transition temperature, T$_C\sim$35\,K.
}
\label{FigChi}
\end{figure}

In Fig.\ref{FigChi}, we present the temperature dependence of the susceptibility data of \rgfao\ below about 5\,K measured in the field cooling (FC) mode with 10\,Oe magnetic field. Inset of Fig.\ref{FigChi} is a plot of the susceptibility data measured in both zero field cooling (ZFC) and FC modes, showing the superconducting transition temperature, T$_C\sim$35\,K \cite{ZCWang2017jpcm}.
The increase of the FC data with decreasing temperature below T$_{mag}\sim$3\,K signals the onset of magnetic ordering of the Gd$^{3+}$ moments. The upturn is not likely due to the small amount of Gd$_2$O$_3$ impurity phases since cubic Gd$_2$O$_3$ exhibits an antiferromagnetic transition at 1.6\,K and monoclinic Gd$_2$O$_3$ has an antiferromagnetic transition at 3.4\,K \cite{prb.58.3212}. This conclusion is consistent with the low temperature upturn of the specific heat data \cite{ZCWang2017jpcm} which suggests a bulk effect of this transition. As for the nature of this magnetic transition, it is most likely to be canted antiferromagnetism. This is consistent with the negative Curie-Weiss temperature obtained from the high temperature magnetic susceptibility data (not shown). Note that for a similar compound \rdfao\ with Dy$^{3+}$ magnetic moments becomes antiferromagnetically ordered below about 10\,K \cite{cm.29.1805}. Of course, future neutron diffraction measurements are needed to clarify this point.

\begin{figure}
\centering
\includegraphics[width=0.9\columnwidth,clip=true]{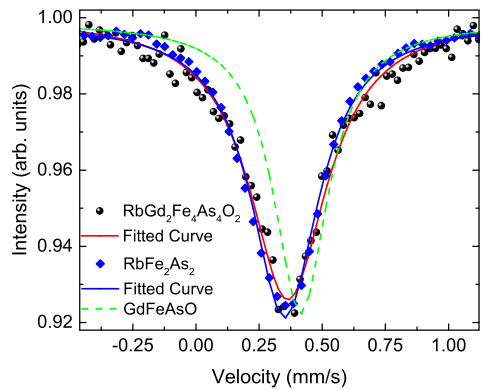}
\caption{
(color online)
The 300\,K $^{57}$Fe M\"ossbauer spectra of \rgfao\ (black dots) and \rfa\ (blue diamond) together with singlet fits. The green dashed line is calculated curve of \gfao\ using hyperfine parameters taken from ref. \cite{PWang2010jpcm}. Note that these spectra were shifted and normalized together for better view.
}
\label{Figcompare}
\end{figure}

One interesting feature of the compound \rgfao\ is self doping induced superconductivity with a transition temperature as high as 35\,K \cite{ZCWang2017jpcm}. As shown by the structural characterization, actual self doping was suggested by the charge redistribution which was reflected from the structural reconstruction, namely the '\gfao' building block becomes slender while the '\rfa' block goes the opposite when compared with the two separate compounds \gfao\ and \rfa\ \cite{ZCWang2017jpcm}. In order to see if this self doping effect really happens locally, namely at the Fe site, we compare the 300\,K MS of the \rgfao\ superconductor with that of the \rfa\ and \gfao\ in Fig.\ref{Figcompare}. The MS of the \gfao\ shown in Fig.\ref{Figcompare} was calculated using hyperfine parameters taken from reference \cite{PWang2010jpcm}. One can see that the center or IS of the MS shifts to positive direction from sample \rfa\ to \rgfao\ and \gfao, reflecting a decrease of the $s$-electron density at the Fe nucleus. To be quantitative, the fitted IS value for sample \rgfao\ is IS\,=\,0.364(6)\,mm/s which is much closer to the value of IS\,=\,0.352(2)\,mm/s for sample \rfa\ when compared to the value of IS\,=\,0.415(4)\,mm/s for sample \gfao\ \cite{PWang2010jpcm}. This is different with the case for the intermediate value of IS\,=\,0.372(2)\,mm/s for the intergrown sample \ckfa\ in comparison with that of samples \kfa\ IS\,=\,0.311(1)\,mm/s and \cfa\ IS\,=\,0.430(2)\,mm/s \cite{SLBudko2017arXiv}. This could be due to different chemical environments of the FeAs neighboring layers, namely the Rb and GdO layers, or alternatively, due to the different lattice dynamics for sample \rgfao\ in comparison with samples \rfa\ and \gfao. To resolve this issue, we made low temperature measurements and found out that, at 5.9\,K, the IS value of sample \rgfao\ (0.495(6)\,mm/s) is right in the middle of that for samples \rfa\ (0.451(4)\,mm/s) and \gfao\ (0.542(5)\,mm/s at 1.9\,K) \cite{PWang2010jpcm} indicating that the lattice dynamic plays a more important role. These results provide direct evidence of the local electronic change which answers the question that the self doping effect does happen on the local scale at the iron nucleus which is also similar to the systematic change of IS with hole doping for the \bkfa\ system \cite{DJohrendt2009pc}.

\begin{figure}
\centering
\includegraphics[width=1\columnwidth,clip=true]{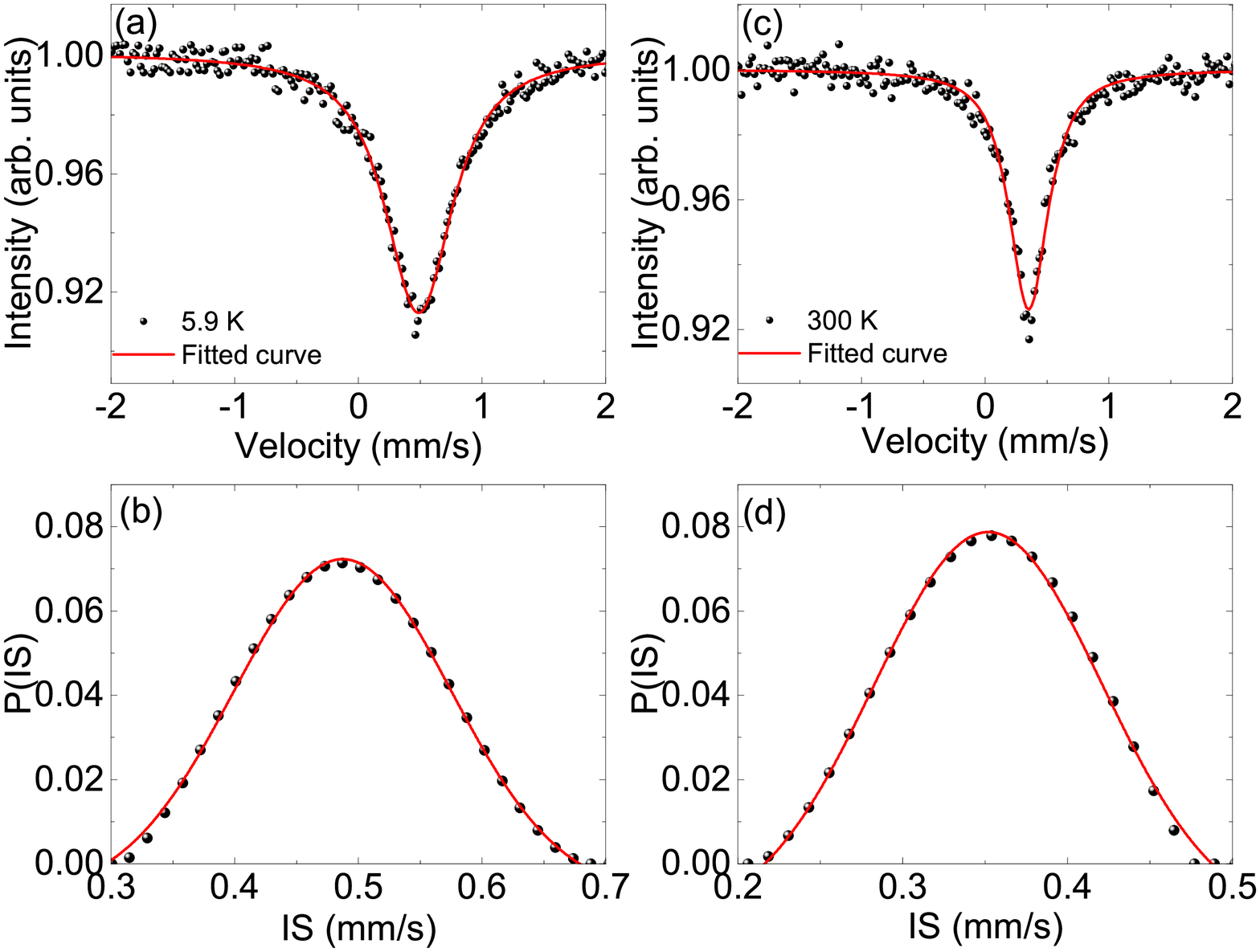}
\caption{
(color online)
$^{57}$Fe M\"ossbauer spectra of \rgfao\ and the corresponding IS distribution measured at 5.9\,K for (a) and (b) and at 300\,K for (c) and (d).
}
\label{Figdist}
\end{figure}

\begin{figure}
\centering
\includegraphics[width=0.8\columnwidth,clip=true]{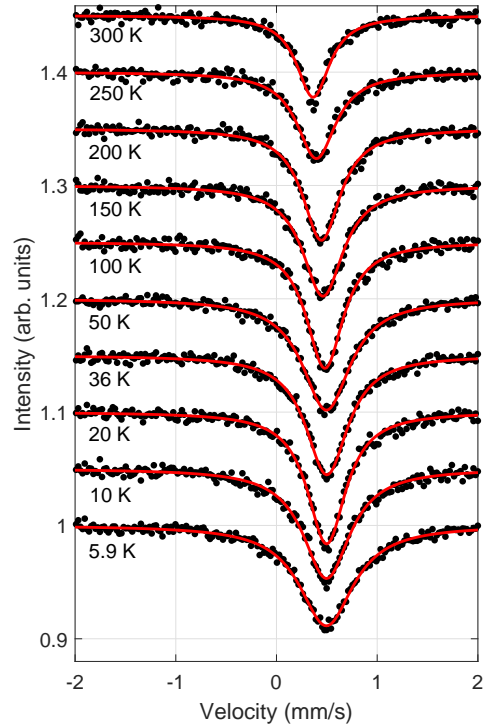}
\caption{
(color online)
$^{57}$Fe M\"ossbauer spectra (black dots) of \rgfao\ measured at indicated temperatures. The red solid lines are singlet fits to the experimental data.
}
\label{Figmoss}
\end{figure}

Another important result is that the MS of \rgfao\ exhibit a singlet pattern which persists to the lowest temperature of 5.9\,K in the current experiment indicating the absence of static magnetic order on the Fe site which is consistent with the superconducting ground state \cite{ZCWang2017jpcm}. Attempts to fit our MS with a distribution of IS have been made as shown in Fig.\ref{Figdist} for the 5.9\,K and 300\,K case. Clearly, the IS distribution can be modeled with one symmetric Guassian profile which is consistent with the fact that there is only one crystallographic site for the Fe ion \cite{ZCWang2017jpcm}. Additionally, trying to fit the spectra with a doublet profile always yield zero quadrupole splitting values. Therefore, we fitted our MS with only one singlet profile as shown in Fig.\ref{Figmoss}, from which reasonable fits can be seen. Clearly, this proves again that our sample is free of impurity phases that contain Fe element or their amounts is too small to be detected by M\"ossbauer spectroscopy \cite{INowik2008jsnm}. The fitted IS and LW are listed in Table \ref{table} together with that of samples \rfa\ and \gfao\ for comparison.

\begin{table}[ht]
\centering
\caption{Hyperfine parameters of \rgfao. IS, isomer shift; LW, spectral full linewidth at half maximum. The 300\,K data for sample \rfa\ and \gfao\ \cite{PWang2010jpcm} were also shown for comparison.}
\label{table} {
\begin{tabular}{c c c c}
\hline \hline
Sample & Temp. (K)  &    IS (mm/s)   &  LW (mm/s)       \\
\hline
	& 5.9	& 0.495(6)	& 0.65(2)	    \\
	& 8.3	& 0.496(6)	& 0.57(2)     \\
	& 10	& 0.495(5)	& 0.52(2)	  \\
	& 15	& 0.500(5)	& 0.44(2)    \\
	& 20	& 0.497(4)	& 0.46(1)	   \\
	& 25	& 0.499(5)	& 0.46(1)	   \\
	& 36	& 0.500(5)	& 0.51(2)	    \\
	& 42	& 0.496(5)	& 0.50(2)	    \\
\rgfao	& 50	& 0.495(5)	& 0.57(2)	   \\
	& 75	& 0.490(5)	& 0.53(2)	  \\
	& 100	& 0.481(5)	& 0.45(2)      \\
	& 125	& 0.467(4)	& 0.40(1)	    \\
	& 150	& 0.461(5)	& 0.46(2)	  \\
	& 200	& 0.446(5)	& 0.45(2)	    \\
	& 225	& 0.432(5)	& 0.43(1)	   \\
	& 250	& 0.396(7)	& 0.46(2)	   \\
	& 300	& 0.364(6)	& 0.40(2)	   \\
\rfa	& 300	& 0.352(2)	& 0.327(7)	   \\
        & 5.9	& 0.451(4)	& 0.40(1)	   \\
\gfao\ \cite{PWang2010jpcm}	& 300	& 0.415(4)	& -	   \\
                            & 1.9	& 0.542(5)	& -	   \\
\hline \hline
\end{tabular}}
\end{table}

\begin{figure}
\centering
\includegraphics[width=0.9\columnwidth,clip=true]{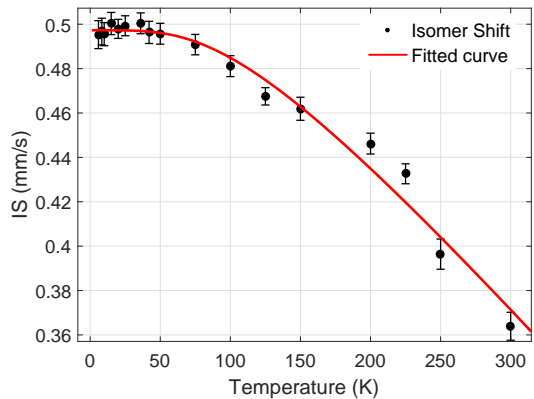}
\caption{
(color online)
Temperature dependence of the IS of \rgfao\ determined from the fits shown in Fig.\ref{Figmoss}. The solid line is theoretical fit to the Debye model as explained in the text.
}
\label{FigIS}
\end{figure}

The temperature dependence of IS, determined from the fits of the MS shown in Fig.\ref{Figmoss}, is shown in Fig.\ref{FigIS} together with theoretical fit using equation (\ref{eqIS}). One can see that IS increases gradually with decreasing temperature and saturates at low temperatures which is the typical behavior of the Debye model. In the Debye approximation of the lattice vibrations, IS can be fitted by the following equation \cite{Mossbook}
\begin{equation}
\begin{split}
IS(T) = IS(0) - \frac{9}{2}\frac{k_BT}{Mc}(\frac{T}{\Theta_D})^3\int_0^{\Theta_D/T}\frac{x^3dx}{e^x-1}
\end{split}
 \label{eqIS}
\end{equation}
where IS(0)=0.497(4)\,mm/s is the temperature independent chemical shift, and the second part is the temperature dependent second order Doppler shift. $k_B$ is the Boltzmann constant, $M$ is the mass of the M\"ossbauer nucleus, $c$ is the speed of light and $\Theta_D$ is the corresponding Debye temperature. A Debye temperature of $\Theta_D=415\pm47$\,K was obtained from the above fit. Similar values of $\Theta_D=409\pm4$\,K \cite{PWang2010jpcm} for \gfao\ and $\Theta_D=474\pm20$\,K \cite{SLBudko2017arXiv} for \kfa\ were found by the same method. We note that the higher uncertainty in our fit mainly comes from the deviation of the experimental data from the theory above $\sim$150\,K coinciding with the broad humpback in the resistivity measurement, which might be related to the incoherent-to-coherent crossover in relation with an emergent Kondo lattice effect \cite{ZCWang2017jpcm,YPWu2016prl}. Certainly, in order to confirm this point, more measurements on other similar compounds with different rare earth elements are under way.

\begin{figure}
\centering
\includegraphics[width=1\columnwidth,clip=true]{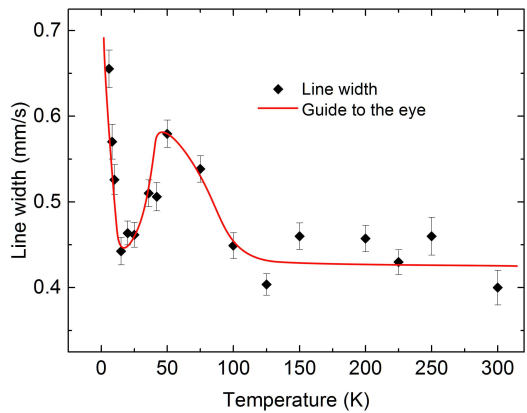}
\caption{
(color online)
Temperature dependence of the spectral LW of \rgfao. The solid lines are guides to the eye.
}
\label{FigLW}
\end{figure}

Another interesting feature of the \rgfao\ compound is the possible interplay of magnetism with superconductivity at low temperatures. Note the magnetic transition temperature of the Gd$^{3+}$ moments, T$_N\sim$3\,K, as shown in Fig.\ref{FigChi}. To see if there is any transferred magnetic fluctuations at the Fe site from the Gd$^{3+}$ moments, we plot the LW as a function of temperature in Fig.\ref{FigLW}. From the first measurement shown in Fig.\ref{FigLW}, we can see that above 100\,K a featureless LW with values in the range of 0.4\,mm/s - 0.46\,mm/s was observed. With decreasing temperature, two LW broadenings can be seen. Broadening of the LW usually reflects the distribution of the surrounding electromagnetic environments experienced by the $^{57}$Fe nucleus.
First, LW broadening usually come from defects or doping induced inhomogeneities. However, these kind of inhomogeneities usually do not change with temperature at relatively low temperatures as in our measurements. Second, some kind of chemical alterations that exist at the sample surface may also give such LW broadenings. However, one would expect a monotonic temperature behavior of the LW broadening, which is inconsistent with our observations that the spectral LW first increase, then decrease and increase again with decreasing temperature. Therefore, these reasons are likely not responsible for our observed LW broadenings here.
Additionally, LW broadening can be induced by magnetic fluctuations as observed in other iron based superconductors \cite{XMMa2014epl,XMMa2013jpcm}. Therefore, we attribute the broadening below about 15\,K to the transferred magnetic fluctuations from the Gd$^{3+}$ moments which orders below about 3\,K as discussed earlier. More interestingly, the broadening below 100\,K, considerably lower than the spin density wave ordering transition temperature for \gfao\ $\sim$128\,K \cite{YLuo2009prb}, might be attributed to magnetic fluctuations on the Fe site. This is consistent with the fact that this LW broadening was completely suppressed within the superconducting state which can be understand as a result of the competition between magnetism and superconductivity in the compound \cite{MDLumsden2010jpcm,JMunevar2011arXiv,DKPratt2009prl}. This suggests that the LW broadenings observed here are induced by intrinsic electronic inhomogeneities arise from competitions between multiple order parameters other than chemical disorders as mentioned above.
It has been widely discussed, in the passed, regarding the interplay between magnetism and superconductivity in the iron based superconductors \cite{PDai2012np,MDLumsden2010jpcm}. And, even after almost ten years of intense study, a final understanding of this issue is still far from reached. Our work provides a new direction to further investigate this problem. Especially, inelastic neutron scattering study of the dynamical magnetic properties would be very important.

\section{Summary}

In summary, $^{57}$Fe M\"ossbauer spectroscopy measurements on \rgfao\ has been performed. The singlet pattern of the spectra indicates that the sample does not exhibit static magnetic order down to 5.9\,K on the Fe sublattice. The observed intermediate value of the isomer shift for our sample \rgfao\ compared to that of the \rfa\ and \gfao, suggests an effective self hole doping at the Fe site. A Debye temperature of $415\pm47$\,K was determined by the temperature dependence of the isomer shift. Most importantly, at the Fe site, we found transferred magnetic fluctuations from the Gd$^{3+}$ moments below 15\,K and magnetic fluctuations of the Fe moments below 100\,K. Our results call for future investigations to study the possible interplay between magnetism and superconductivity in this iron-based superconductor.

\section{Acknowledgements}

This work was supported by the Fundamental Research Funds for the Central Universities under Grant number 223000/862616 and the National Key Research and Development Program of China under Grant number 2016YFA0300202.

\bibliography{rgfao}

\end{document}